\nonstopmode

\newcommand{\ie}{i.e.{}}
\newcommand{\eg}{e.g.{}}
\newcommand{\eV}{\U{eV}}
\newcommand{\U}[1]{\,{\rm{#1}}}
\newcommand{\X}[1]{_{\mathrm{#1}}}
\newcommand{\Sum}{\sum\limits}
\newcommand{\differential}{\>\mathrm d}
\newcommand{\XUV}{\textsc{xuv}}
\newcommand{\nbh}{\hbox{-}}
\newcommand{\Xray}{X\nbh{}ray}
\newcommand{\xray}{x\nbh{}ray}
\newcommand{\atopa}[2]{\stackrel{#1}{#2}}

\documentclass[12pt]{iopart}
\usepackage{iopams,bbm,CJK,datetime,fancyhdr,color,subeqnarray,cite,graphicx}

\usepackage[nativepdf,bookmarks,bookmarksopen,bookmarksnumbered,raiselinks,%
pdftitle={Rate equations for nitrogen molecules in ultrashort and intense x-ray pulses},%
pdfauthor={Ji-Cai Liu (刘纪彩), Nora Berrah, Lorenz S. Cederbaum,
James P. Cryan, James M. Glownia, Kenneth J. Schafer, Christian Buth},%
pdfsubject={Molecular Physics},%
pdfkeywords={X rays, photoabsorption, Auger decay, light-matter interaction,
nitrogen molecule, multiply-charged atoms, molecular fragmentation,
metastable dication, frustrated absorption, molecular rate equations,
ultrafast, intense}]{hyperref}

\begin{document}
\title[Rate equations for nitrogen molecules]{Rate equations for nitrogen
molecules in ultrashort and intense \xray~pulses\footnote{%
This is an author-created, un-copyedited version of an article accepted
for publication in Journal of Physics B: Atomic, Molecular and Optical Physics.
IOP Publishing Ltd is not responsible for any errors or omissions in
this version of the manuscript or any version derived from it.
The Version of Record is available online at
\href{http://dx.doi.org/10.1088/0953-4075/49/7/075602}
                   {doi:10.1088/0953-4075/49/7/075602}.}
}

\begin{CJK*}{UTF8}{}
\author{Ji-Cai Liu ({\CJKfamily{gbsn}刘纪彩}),$^{1,2}$ Nora Berrah,$^{3}$
Lorenz S.{} Cederbaum,$^4$ James P.{} Cryan,$^{5,6}$ James M.{} Glownia$,^{5,6}$
Kenneth J.{} Schafer,$^{5,7}$ and Christian Buth$^{2,4,5,7}$}
\end{CJK*}

\address{$^1$ Department of Mathematics and Physics, North China Electric Power
University, 102206~Beijing, China}
\address{$^2$ Max-Planck-Institut f\"ur Kernphysik, Saupfercheckweg~1,
69117~Heidelberg, Germany}
\address{$^3$ Department of Physics, University of Connecticut,
2152 Hillside Road, U-3046, Storrs, Connecticut~06269, USA}
\address{$^4$ Theoretische Chemie, Physikalisch-Chemisches Institut,
Im Neuenheimer Feld~229, Ruprecht-Karls-Universit\"at Heidelberg,
69120~Heidelberg, Germany}
\address{$^5$ The PULSE Institute for Ultrafast Energy Science, SLAC
National Accelerator Laboratory, Menlo Park, California~94025, USA}
\address{$^6$ Department of Physics, Stanford University, Stanford,
California~94305, USA}
\address{$^7$ Department of Physics and Astronomy, Louisiana State
University, Baton Rouge, Louisiana~70803, USA}

\eads{\mailto{christian.buth@web.de}, World Wide Web:
\texttt{\href{http://www.christianbuth.name}{www.christianbuth.name}}}

\begin{abstract}
We study theoretically the quantum dynamics of nitrogen molecules~(N$_2$)
exposed to intense and ultrafast x-rays at a wavelength of~$1.1 \U{nm}$
($1100 \eV$ photon energy) from the Linac Coherent Light Source~(LCLS)
free electron laser.
Molecular rate equations are derived to describe the intertwined
photoionization, decay, and dissociation processes occurring for~N$_2$.
This model complements our earlier phenomenological approaches, the single-atom,
symmetric-sharing, and fragmentation-matrix models of J.{} Chem.{}
Phys.{} \textbf{136}, 214310 (2012).
Our rate-equations are used to obtain the effective pulse energy
at the sample and the time scale for the dissociation
of the metastable dication~N$_2^{2+}$.
This leads to a very good agreement between the theoretically and
experimentally determined ion yields and, consequently, the average
charge states.
The effective pulse energy is found to decrease with shortening
pulse duration.
This variation together with a change in the molecular fragmentation pattern and
frustrated absorption---an effect that reduces absorption of x-rays
due to (double) core hole formation---are the causes for the drop of the average
charge state with shortening LCLS pulse duration discovered previously.
\end{abstract}

%
%
%
%
%
%



\noindent{\textit{Date}\/}: \leftline{16 March 2017}

\definecolor{mablack}{rgb}{0,0,0}
\definecolor{mared}{rgb}{1,0,0}
\definecolor{magreen}{rgb}{0,1,0}
\definecolor{mablue}{rgb}{0,0,1}
\definecolor{mamagenta}{rgb}{1,0,1}

\section{Introduction}

With the development of \xray~lasers, especially the state-of-the-art
\xray~free electron lasers~(FELs), the exploration of the interaction
of intense and ultrafast x-rays with matter has become a scientific
frontier~\cite{Hoener:FA-10,Fang:DC-10,Cryan:AE-10,Young:FE-10,%
Vartanyants:CP-11,Bernitt:LO-12,Glover:XO-12,Boutet:SC-12,%
Rohringer:AI-12,Salen:EV-12,Thomas:XC-12,Starodub:DP-12,Murphy:MP-12,%
Petrovic:XF-12,Gorkhover:NP-12,Schorb:SD-12,Redecke:NI-13,Schulz:FA-15}.
The extreme pulse characteristics of operational \xray~FELs---such as the
the Linac Coherent Light Source~(LCLS)~\cite{LCLS:CDR-02,Emma:FL-10}
in Menlo Park, California, USA---are ultrahigh brightness, femtosecond
pulse duration, short wavelengths, and transverse coherence.
The novel properties of x-rays enable the study of ultrafast
multiphoton interaction of x-rays with matter and spawned the subfield of
\xray~quantum optics~\cite{Adams:QO-13}.
Specifically, the intense x-rays from FELs induce two-photon
absorption~\cite{Doumy:NA-11} (extreme
ultraviolet~(\XUV)~\cite{Nabekawa:DH-05}),
Rabi flopping~\cite{Kanter:MA-11,Cavaletto:RF-12},
photoelectron holography~\cite{Krasniqi:IM-10},
single-pulse pump-probe experiments~\cite{Fang:MI-12},
stimulated \xray~Raman scattering~\cite{Sun:PX-10}, and
four-wave mixing~\cite{Sun:PX-10} (\XUV{}~\cite{Bencivenga:FW-15}), suppress
the branching ratio of Auger decay~\cite{Liu:AE-10} or
produce interference effects in Auger decay~\cite{Demekhin:IE-11,Buth:RM-15}.
Saturable~\cite{Nagler:TA-09} and frustrated
absorption~\cite{Hoener:FA-10,Young:FE-10,Buth:UA-12}
(\XUV{}~\cite{Gerken:TM-14}) reduce radiation damage of the sample
and may thus be beneficial for diffraction experiments at
FELs~\cite{Neutze:BI-00}.
The combination of intense and ultrafast x-rays with an optical laser
offers perspectives for the control of
\xray~lasing~\cite{Darvasi:OC-14}, high-order harmonic generation
in the kiloelectronvolt regime~\cite{Buth:NL-11,Kohler:EC-12,Buth:KE-13,%
Buth:HO-15}, and the production of high-energy frequency
combs~\cite{Cavaletto:FC-13,Cavaletto:HF-14}.

Initially, atoms were studied in intense and ultrafast soft \xray~pulses from
LCLS~\cite{Young:FE-10}.
Neon was chosen as a prototypical atom~\cite{Young:FE-10} and it
was found that, for sufficiently high photon energies, all atomic electrons
are ionized by one-\xray-photon absorption and atoms even become transparent
at high \xray~intensity due to rapid ejection of inner-shell electrons.
Ionization by x-rays~\cite{Young:FE-10} (\XUV~\cite{Sorokin:PE-07,Makris:MM-09})
progresses \emph{from the inside out} in contrast to strong-field
ionization with intense optical lasers where ionization of the
outermost electrons occurs, \ie, there ionization proceeds
\emph{from the outside in}.
The quantum dynamics of neon atoms in intense and ultrashort \xray~radiation
was successfully described theoretically with a rate-equation
model~\cite{Rohringer:XR-07,Young:FE-10} foreshadowing extension to
more complex systems.

Understanding the response of molecules to intense and ultrafast
x-rays poses substantial complications over the description of
atoms~\cite{Hoener:FA-10,Cryan:AE-10,Fang:DC-10,Buth:UA-12,%
Fang:MI-12,Cryan:AE-12};
namely, in addition to photoionization and intraatomic decay
processes, for molecules, there are also molecular fragmentation,
and sharing of the charges in the valence shells among the
nuclei which need to be regarded.
The quantum dynamics induced in molecules by soft x-rays~\cite{Hoener:FA-10,%
Cryan:AE-10,Fang:DC-10,Buth:UA-12,Fang:MI-12,Cryan:AE-12} from FELs
(\XUV~\cite{Sorokin:MP-06,Sato:DT-08,Jiang:FP-09}),
has been studied in experiments during the last few years.
The first ever study of a molecule, namely nitrogen~(N$_2$), in intense and
ultrafast x-rays was performed at LCLS~\cite{Hoener:FA-10}
where N$_2$ was chosen because it is an important but
simple molecule.
The experiment used x-rays with a wavelength of~$1.1 \U{nm}$, \ie,
$1100 \eV$ photon energy.
We found~\cite{Hoener:FA-10} that the absorption of x-rays is frustrated,
\ie, pulses with a comparable energy but decreasing duration lead to
smaller average charge states compared with longer pulses and thus a
reduced absorption of x-rays.
Unlike the \XUV~photon energy regime in which multiphoton
absorption causes single or multiple ionization~\cite{Sorokin:MP-06,Sato:DT-08,%
Jiang:FP-09}, predominantly only one-photon ionization occurs in the
\xray~regime~\cite{Young:FE-10,Hoener:FA-10,Buth:UA-12,Fang:MI-12} which
is a substantial simplification of the problem.

To find a theoretical description of~N$_2$ in intense and ultrafast x-rays,
we devised a series of phenomenological models
of increasing sophistication, namely, a single-atom model, a
symmetric-sharing model, and a fragmentation-matrix model~\cite{Buth:UA-12}.
It was found that a single-atom model---which assumes that the molecular
ion yields are the same as the ion yields that are obtained from a single
atom---is not capable to describe the experimental data for~N$_2$
due to a redistribution of valence charges
over both N~atoms in the molecule after ionization and prior breakup.
Therefore, a symmetric-sharing model was tried next in which the
molecular charge is distributed evenly between the two atoms.
The model was found to lead to a similarly deficient description
as the single-atom model.
Thus the insights gained from these two models were used to devise
heuristically a fragmentation-matrix model in which there is only a
partial redistribution of charges between the two atoms.
This model clearly reveals the relevance of
the redistribution of charge and a different weighting of
electronic processes and nuclear dynamics on short and long
time scales for similar nominal pulse energy but varying pulse duration.

In this article, we study sequential multiple ionization
of~N$_2$ induced by intense and ultrafast x-rays from~LCLS
with comparable nominal pulse energy but different pulse durations.
In the theory~\sref{sec:theory}, we extend our modeling
from~\cite{Buth:UA-12} by devising a molecular rate-equation
formalism\footnote{%
The model for~N$_2$ in intense and ultrafast x-rays
of Ref.~\cite{Hoener:FA-10} is also based on molecular rate
equations and was developed by Oleg Kornilov and Oliver Gessner.
The rate equations in this work were devised independently
and their results are consistent with the phenomenological
models of~\cite{Buth:UA-12} and further analysis of the
experimental data in~\cite{Fang:MI-12} unlike the previous attempt
in Ref.~\cite{Hoener:FA-10}.}
to describe the quantum dynamics induced by \xray~absorption.
The molecular rate equations are formulated for neutral~N$_2$ and
molecular cations with single core holes~(SCHs), double core holes~(DCHs)
on a single site~(ssDCH), and double core holes on two
sites~(tsDCH)~\cite{Tashiro:MD-10,Berrah:DC-11,Salen:EV-12,%
Osipov:DC-12,Buth:UA-12}.
Further rate equations for valence ionized molecular configurations and
even triple core holes round off the set.
In total, we consider all possible one-photon absorption and Auger decay
channels for molecular charge states up
to~N$_2^{3+}$~\cite{Buth:UA-12,SuppData}\nocite{Mathematica:pgm-V10.1}.
We treat molecular charge states beyond~N$_2^{3+}$ as fragmented
into single atoms and we take all single atom charge states into account
in terms of atomic rate equations~\cite{Buth:UA-12,SuppData}.
By considering the finite lifetime of the metastable molecular
dication~N$_2^{2+}$ and the sharing of the valence electrons between
the two nuclei of the molecule, the experimental ion yields and the
average charge state are well reproduced in the results and
discussion~\sref{sec:results}.
Our present molecular rate-equation model shows that the
differences in the ion yields observed for pulses of
varying durations are due to the competition between photoabsorption,
Auger decay processes, and molecular fragmentation.
Frustrated absorption is explained as a reduction of photoabsorption
by core hole formation.
The effective pulse energy at the sample and the rate of the dissociation
of the molecular dication are obtained with the molecular rate-equation model
by comparing theoretical and experimental data.
Our rate-equation model represents an improvement over the
fragmentation-matrix model~\cite{Buth:UA-12} as it describes
the quantum dynamics of the involved molecular charge states
and the breakup process.
Nonetheless, the results from our model agree well with our earlier
findings~\cite{Buth:UA-12}.
Conclusions are drawn in~\sref{sec:conclusion}.
All details of the molecular rate-equation model and the calculations
in this article are provided in the Supplementary Data~\cite{SuppData}.

Atomic units~\cite{Hartree:WM-28,Szabo:MQC-89} are used throughout
unless stated otherwise.
For the conversion of a decay width~$\Gamma$ in electronvolts to a
lifetime~$\tau$ in femtoseconds, we use the
relation~$\tau = \frac{\hbar}{\Gamma} = \frac{0.658212 \U{eV \, fs}}{\Gamma}$.

\section{Theory}
\label{sec:theory}

\begin{figure}
  \begin{indented}
    \item[]\includegraphics[clip,width=0.8\hsize]{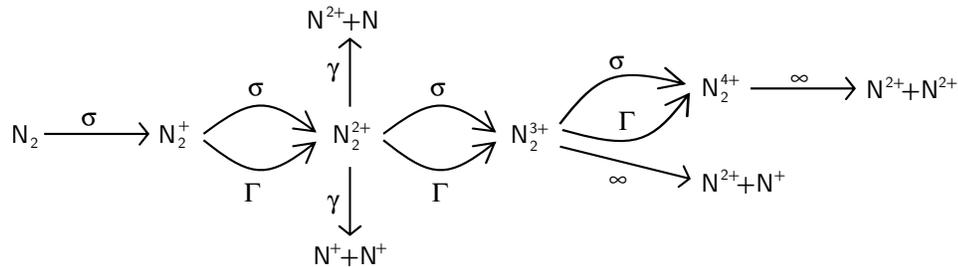}
    \caption{Schematic of the interaction of~N$_2$ with intense x-rays.
             The molecular processes displayed are photoionization by
             one-\xray-absorption~``$\sigma$'', Auger decay of core-ionized
             states ``$\Gamma$'', dissociation of the metastable
             molecular ions~N$_2^{2+}$ with a fragmentation rate ``$\gamma$'',
             and instantaneous fragmentation ``$\infty$''.
             In the last step of the interaction ``$\infty$'' occurs, if no
             core holes are present in~N$_2^{3+}$;
             otherwise ``$\sigma$'' or ``$\Gamma$'' take place.}
    \label{fig:molionH}
  \end{indented}
\end{figure}

We develop a rate-equation description for~N$_2$ in intense and ultrafast x-rays
which includes the molecular processes schematically depicted
in~\fref{fig:molionH}.
Initially, all N$_2$~molecules are in their ground state.
The absorption of an x~ray creates the molecular cation~N$_2^+$ which
is frequently metastable with respect to dissociation.
Thus N$_2^+$~remains a molecule until either another x~ray is absorbed
or, if there is a core hole, Auger decay takes
place~\cite{Siegbahn:FM-69,Stalherm:DI-69,Moddeman:DA-71,Stolte:DP-98}.
Both processes promote~N$_2^+$ to the dication~N$_2^{2+}$.
Even~N$_2^{2+}$ that, in our case, is predominantly
produced by Auger decay, is frequently metastable with a
comparatively long lifetime of~${\sim} 100 \U{fs}$ with respect to
dissociation~\cite{Wetmore:TO-86,Beylerian:CE-04}.
There are three channels for~N$_2^{2+}$ to evolve into;
first, N$_2^{2+}$~fragments causing the end of the molecule;
second, an x~ray is absorbed, and, third, if there is a core hole,
Auger decay occurs.
Both the second and the third channel promote~N$_2^{2+}$ to~N$_2^{3+}$.
If N$_2^{3+}$ has no core holes, we assume that it fragments immediately
into atoms;
otherwise, again \xray~absorption and Auger decay may occur
turning~N$_2^{3+}$ into~N$_2^{4+}$ which is considered to fragment
immediately into two N$^{2+}$~atoms~\cite{Codling:CS-91,Baldit:CE-05}.
From here onwards, only atomic fragments are considered and their further
evolution is governed by atomic rate equations.

Molecular rate equations are formulated for~N$_2$ incorporating the
steps mentioned in the previous paragraph.
%
%
%
%
%
%
%
%
%
%
%
Altogether there are thirty-three molecular configurations from
neutral~N$_2$ to triply charged~N$_2^{3+}$ with different
combinations of the electronic configurations of the two atomic sites.
The quantum dynamics of the atomic fragments after breakup is described with
an atomic rate-equation model which is modified in order to funnel
probability from the molecular rate equations into the atomic rate equations.
All in all, we extend nine of the thirty-six atomic rate equations
to describe the transfer of probability from molecular configurations
to atomic configurations.
The molecular and atomic rate equations incorporate all energetically
accessible atomic one-photon absorption processes together with
fluorescence and Auger decay of multiply-charged
nitrogen atoms~\cite{Buth:UA-12};
simultaneous multiphoton
absorption~\cite{Rohringer:XR-07,Doumy:NA-11,Scheit:NT-16}
and multielectron processes such as shake~off~\cite{Doumy:NA-11}
and double Auger decay~\cite{Aberg:TP-75}, however, are neglected.
In what follows, we write down selected molecular rate equations
to show the guiding principles.
The complete set of equations is provided in the Supplementary
Data~\cite{SuppData}.

\subsection{Neutral nitrogen molecule~$\mathrm{N}_2$}
\label{sec:neutral}

We specify probabilities~$\tilde P_{ijk,\ell mn}(t)$ at time~$t$
for~$0 \leq i, j, \ell, m \leq 2 \, \land \, 0 \leq k, n \leq 3$
to find an N$_2$~molecule in which the individual N~atoms have the electronic
configurations~$1s^i 2s^j 2p^k$ and $1s^{\ell} 2s^m 2p^n$, respectively,
which are abbreviated by~$\{ijk\}$ and $\{\ell mn\}$.
Molecular configurations for~N$_2$ are written succinctly as~$\{ijk, \ell
mn\}$\footnote{%
Note that~$\{ijk, \ell mn\} = \{\ell mn, ijk\}$ holds,
\ie, only distinct molecular configurations up to a permutation
are considered to avoid double counting.}.
\Xray~absorption by~N$_2$ is governed by a single rate equation:
\begin{equation}
  \label{eq:molN2}
  \frac{\differential \tilde P_{223,223}(t)}{\differential t}
    = -2 \, \sigma_{223} \, \tilde P_{223,223}(t) \, J\X{X}(t) \; ,
\end{equation}
with the total one-\xray-photon absorption cross
section~$\sigma_{ijk}$ of an N~atom in configuration~$\{ijk\}$
where~$i = 2$, $j = 2$, $k = 3$~denotes the neutral N~atom in the
ground state.
The factor of~$2$ accounts for the two atoms in~N$_2$ and the minus sign
indicates a depletion of~N$_2$ in the ground state by \xray~absorption.
Further, $J\X{X}(t)$ is the \xray~photon flux at time~$t$.
Note that the \xray~flux has, in principle, both temporal and spatial
dependence due to the longitudinal and transverse profiles of the
\xray~beam~\cite{Buth:UA-12}.
For simplicity of notation, we do not explicitly write spatial coordinates
for the quantities involved but state all equations for a flux
that depends only on time~$J\X{X}(t)$.

\subsection{Molecular nitrogen cation~$\mathrm{N}_2^+$}
\label{sec:cation}

For the singly-charged molecule~N$_2^+$ with a SCH, the governing
rate equation reads
\begin{eqnarray}
  \label{eq:P123,223}
  \frac{\differential \tilde P_{123,223}(t)}{\differential t}
    &=& 2 \, \sigma_{123 \leftarrow 223} \, \tilde P_{223,223}(t)
    \, J\X{X}(t) \\
  &&{} - (\sigma_{123} + \sigma_{223}) \, \tilde P_{123,223}(t) \,
    J\X{X}(t) - \Gamma_{123} \, \tilde P_{123,223}(t) \nonumber \; ,
\end{eqnarray}
where~$\sigma_{\ell mn \leftarrow ijk}$ stands for the one-\xray-photon
absorption cross section which causes an N~atom with the
configuration~$\{ijk\}$ to transition into the configuration~$\{\ell mn\}$.
With~$\Gamma_{ijk}$ we denote the total decay rate of an
N~atom in configuration~$\{ijk\}$ to any final state.
The first term on the right-hand side of~\eref{eq:P123,223} is the
rate of production of~$\{123,223\}$ by \xray~absorption;
the second term describes the rate of depletion by
\xray~absorption;
and the third term represents the rate of decay.

Further molecular rate equations are formulated for the other
cationic configurations of~N$_2^+$ which are~$\{213,223\}$ and $\{222,223\}$;
the resulting molecular rate equations are expressed
analogously to~\eref{eq:P123,223}~\cite{SuppData}.
Only few nondissociated~N$_2^+$ molecules were detected in the
experiment~\cite{Hoener:FA-10} which justifies that we do not
treat them as a channel but break them into atomic
fragments at~$t \to \infty$.

\subsection{Molecular nitrogen dication~$\mathrm{N}_2^{2+}$}
\label{sec:dication}

For the molecular dication~N$_2^{2+}$ with only valence vacancies, we have the
exemplary rate equation
\begin{eqnarray}
  \label{eq:P212,223}
  \frac{\differential \tilde P_{212,223}(t)}{\differential t}
    &=& \sigma_{212 \leftarrow 213} \, \tilde P_{213,223}(t) \, J\X{X}(t)
    + \sigma_{212 \leftarrow 222} \, \tilde P_{222,223}(t) \, J\X{X}(t) \\
  &&{} - (\sigma_{212} + \sigma_{223}) \, \tilde P_{212,223}(t) \, J\X{X}(t)
    + \Gamma_{212 \leftarrow 123} \, \tilde P_{123,223}(t) \nonumber \\
  &&{} - \gamma \, \tilde P_{212,223}(t) \; , \nonumber
\end{eqnarray}
where~$\Gamma_{\ell mn \leftarrow ijk}$ is the decay rate of the atom in
configuration~$\{ijk\}$ into~$\{\ell mn\}$.
The fragmentation rate of the valence-ionized molecular dication~N$_2^{2+}$ is
indicated by~$\gamma$ which is presumed not to depend on the molecular
configuration.
We assume that all~N$_2^{2+}$ eventually dissociate because only
few nondissociated~N$_2^{2+}$ molecules were detected in
the experiment~\cite{Hoener:FA-10}.

Since~N$_2^{2+}$ dissociates either symmetrically
into~$\mathrm{N}^+ + \mathrm{N}^+$ or asymmetrically
into~$\mathrm{N} + \mathrm{N}^{2+}$, the
parameters~$f_{\mathrm{N}^+ + \mathrm{N}^+}$ and
$f_{\mathrm{N} + \mathrm{N}^{2+}}$ are introduced~\cite{Buth:UA-12} to account
for the probabilities to dissociate into the fragmentation channels
\begin{subeqnarray}
  \label{eq:fN2+}
  \slabel{eq:fN2+sym}
  &&\mathrm{N}_2^{2+} \stackrel{f_{\mathrm{N}^+ + \mathrm{N}^+}}
    {\longrightarrow} \mathrm{N}^+ + \mathrm{N}^+ \; , \\
  \slabel{eq:fN2+asy}
  &&\mathrm{N}_2^{2+} \stackrel{f_{\mathrm{N} + \mathrm{N}^{2+}}}
    {\longrightarrow} \, \mathrm{N}^{\phantom{+}} + \mathrm{N}^{2+} \; ,
\end{subeqnarray}
with~$f_{\mathrm{N}^+ + \mathrm{N}^+} + f_{\mathrm{N} + \mathrm{N}^{2+}} = 1$.
The molecular probability of~N$_2^{2+}$ upon breakup needs to be funneled
into the atomic rate equations.
For example, the atomic rate equation of a neutral N~atom has to be modified to
\begin{eqnarray}
  \label{eq:P223}
  \frac{\differential P_{223}(t)}{\differential t} &=& -\sigma_{223} \,
    P_{223}(t) \, J\X{X}(t) \nonumber \\
  &&{} + f_{\mathrm{N} + \mathrm{N}^{2+}} \; \frac{\gamma}{2} \; \Bigl[
    \tilde P_{203,223}(t) + \tilde P_{212,223}(t)
    + \tilde P_{221,223}(t) \\
  &&\qquad\qquad\ \ {} + \tilde P_{213,213}(t) + \tilde P_{213,222}(t)
    + \tilde P_{222,222}(t) \Bigr] \; , \nonumber
\end{eqnarray}
where~$P_{\ell mn}(t)$ is the probability to find an N~atom
at time~$t$ in electronic configuration~$\{\ell mn\}$.
The last term on the right-hand side of~\eref{eq:P223} is new
compared with the original atomic rate equation and it is responsible for
funneling probability of neutral N~atoms due to molecular fragmentation
from the molecular configurations~$\{203,223\}$, $\{212,223\}$,
$\{221,223\}$, $\{213,213\}$, $\{213,222\}$, and $\{222,222\}$.
Similar modifications are made to the atomic rate equations
for the configurations~$\{213\}$, $\{222\}$, $\{203\}$, $\{212\}$,
$\{221\}$, $\{113\}$, $\{122\}$, and $\{023\}$~\cite{SuppData}.
Other atomic rate equations are not altered.

For a valence hole and a SCH on one atom, the dynamics is quantified by
the molecular rate equation
\begin{eqnarray}
  \frac{\differential \tilde P_{113,223}(t)}{\differential t}
    &=& \sigma_{113 \leftarrow 213} \, \tilde P_{213,223}(t) \, J\X{X}(t)
    + \sigma_{113 \leftarrow 123} \, \tilde P_{123,223}(t) \, J\X{X}(t) \\
  &&{} - (\sigma_{113} + \sigma_{223}) \tilde P_{113,223}(t)
    \, J\X{X}(t) - \Gamma_{113} \, \tilde P_{113,223}(t) \; , \nonumber
\end{eqnarray}
for a tsDCH by
\begin{eqnarray}
  \frac{\differential \tilde P_{123,123}(t)}{\differential t}
    &=& \sigma_{123 \leftarrow 223} \, \tilde P_{123,223}(t) \, J\X{X}(t)
    - 2 \, \sigma_{123} \, \tilde P_{123,123}(t) \,J\X{X}(t) \\
  &&{} - 2 \, \Gamma_{123} \, \tilde P_{123,123}(t) \; , \nonumber
\end{eqnarray}
and for a ssDCH by
\begin{eqnarray}
  \frac{\differential \tilde P_{023,223}(t)}{\differential t}
    &=& \sigma_{023 \leftarrow 123} \, \tilde P_{123,223}(t) \, J\X{X}(t)
    - (\sigma_{023} + \sigma_{223}) \tilde P_{023,223}(t) \, J\X{X}(t) \\
  &&{} - \Gamma_{023} \, \tilde P_{023,223}(t) \; . \nonumber
\end{eqnarray}
We see from these expressions, that molecular fragmentation is not
included when a SCH or DCH is present because the time scale
is, in this case, determined by the decay time of the vacancies
which is much shorter than the fragmentation time.

The other dicationic configurations of~N$_2^{2+}$ are~$\{213,213\}$,
$\{222,222\}$, $\{123,213\}$, $\{123,222\}$, $\{213,222\}$, $\{203,223\}$,
$\{221,223\}$, and $\{122,223\}$;
the involved molecular rate equations are formulated in analogy to the
above scheme~\cite{SuppData}.

\subsection{Molecular nitrogen trication~$\mathrm{N}_2^{3+}$}
\label{sec:trication}

Further photoionization or Auger decay of the molecular dication~N$_2^{2+}$
produces the trication~N$_2^{3+}$.
No bound states were found for~N$_2^{3+}$ ions in the midst of
many Coulomb repulsive potentials~\cite{Bandrauk:ES-99}.
Therefore, we assume that those N$_2^{3+}$ ions with no core
holes fragment into~N$^{2+}$ and $\mathrm N^+$ at the instant
of their formation by Auger decay or photoionization.
The probability for the formation of~N$_2^{3+}$ ions without core
holes is small due to small valence ionization cross sections at the
\xray~photon energy~\cite{Als-Nielsen:EM-01}.
In contrast, N$_2^{3+}$~with core holes is still treated as a molecule,
\ie, no immediate fragmentation is assumed;
the~N$_2^{3+}$ are only broken up by Auger decay or photoionization because
of the faster time scale of Auger decay compared with molecular fragmentation.

For the triply charged molecular configurations, with a SCH and two
valence holes, we have the exemplary rate equations
\begin{eqnarray}
  \frac{\differential \tilde P_{121,223}(t)}{\differential t}
    &=& \sigma_{121 \leftarrow 221} \, \tilde P_{221,223}(t)
    \, J\X{X}(t) + \sigma_{121 \leftarrow 122} \, \tilde P_{122,223}(t)
    \, J\X{X}(t) \nonumber \\
  &&{} - (\sigma_{121} + \sigma_{223}) \, \tilde P_{121,223}(t) \,
    J\X{X}(t) + \Gamma_{121 \leftarrow 023} \, \tilde P_{023,223}(t) \\
  &&{} - \Gamma_{121} \, \tilde P_{121,223}(t) \; , \nonumber \\
  \frac{\differential \tilde P_{212,123}(t)}{\differential t}
    &=& \sigma_{123 \leftarrow 223} \, \tilde P_{212,223}(t) \, J\X{X}(t)
    + \sigma_{212 \leftarrow 213} \, \tilde P_{123,213}(t) \, J\X{X}(t) \\
  &&{} + \sigma_{212 \leftarrow 222} \, \tilde P_{123,222}(t) \,
    J\X{X}(t) - (\sigma_{212} + \sigma_{123}) \, \tilde P_{212,123}(t)
    \, J\X{X}(t) \nonumber\\
  &&{} + 2 \, \Gamma_{212 \leftarrow 123} \,
    \tilde P_{123,123}(t) - \Gamma_{123} \, \tilde P_{212,123}(t) \; ,
    \nonumber
\end{eqnarray}
a tsDCH and a valence hole
\begin{eqnarray}
  \frac{\differential \tilde P_{113,123}(t)}{\differential t}
    &=& \sigma_{113 \leftarrow 213} \, \tilde P_{123,213}(t) \, J\X{X}(t)
    + \sigma_{123 \leftarrow 223} \, \tilde P_{113,223}(t) \, J\X{X}(t)
    \nonumber \\
  &&{} + 2 \, \sigma_{113 \leftarrow 123} \, \tilde P_{123,123}(t) \,
    J\X{X}(t) - (\sigma_{113} + \sigma_{123})\\
  &&{} \times \tilde P_{113,123}(t) \, J\X{X}(t) - (\Gamma_{113}
    + \Gamma_{123}) \, \tilde P_{113,123}(t) \; , \nonumber
\end{eqnarray}
a ssDCH and a valence hole
\begin{eqnarray}
  \frac{\differential \tilde P_{013,223}(t)}{\differential t}
    &=& \sigma_{013 \leftarrow 113} \, \tilde P_{113,223}(t) \,
    J\X{X}(t) + \sigma_{013 \leftarrow 023} \, \tilde P_{023,223}(t)
    \, J\X{X}(t)\\
  &&{} - (\sigma_{013} + \sigma_{223}) \, \tilde P_{013,223}(t) \, J\X{X}(t)
    - \Gamma_{013} \, \tilde P_{013,223}(t) \; , \nonumber
\end{eqnarray}
and a ssDCH and a SCH
\begin{eqnarray}
  \frac{\differential \tilde P_{023,123}(t)}{\differential t}
    &=& 2 \, \sigma_{023 \leftarrow 123} \, \tilde P_{123,123}(t)
    \, J\X{X}(t) + \sigma_{123 \leftarrow 223} \, \tilde P_{023,223}(t)
    \, J\X{X}(t) \\
  &&{} - (\sigma_{023} + \sigma_{123}) \, \tilde P_{023,123}(t) \, J\X{X}(t)
    - (\Gamma_{023} + \Gamma_{123}) \, \tilde P_{023,123}(t) \; . \nonumber
\end{eqnarray}

The other tricationic configurations of~N$_2^{3+}$ are~$\{113,213\}$,
$\{113,222\}$, $\{122,213\}$, $\{023,213\}$, $\{203,123\}$, $\{112,223\}$,
$\{103,223\}$, $\{122,222\}$, $\{023,222\}$, $\{221,123\}$, $\{122,123\}$,
and $\{022,223\}$;
the resulting molecular rate equations are written accordingly~\cite{SuppData}.

\subsection{Higher charge states of a nitrogen molecule}

Molecular effects are centered around the steps described
in~\sref{sec:neutral}, \ref{sec:cation}, \ref{sec:dication},
and \ref{sec:trication}~\cite{Buth:UA-12}.
For even higher charge states, the interaction is basically
those of independent atoms and no molecular rate equations are formulated.
Instead, immediate fragmentation is assumed for the molecular
tetracation~N$_2^{4+}$ where the four charges are shared equally between
the two nitrogen nuclei~\cite{Codling:CS-91,Baldit:CE-05}.
The probability of the fragments is funneled into the atomic
rate equations which govern the time evolution from this point
onwards~\cite{SuppData}.

\section{Results and discussion}
\label{sec:results}

To predict the ion yields and the average charge state of~N$_2$ in
intense and ultrafast x-rays, we use a number of
computational parameters which were either taken from
experiments~\cite{Stolte:DP-98,Hoener:FA-10} or calculated~\cite{Buth:UA-12}.
The \xray~field parameters used here are listed in~\tref{tab:parameters}
and are the same as the ones in~\cite{Hoener:FA-10,Buth:UA-12}.
The expression for the beam profile of LCLS pulses can be
found in~\cite{Buth:UA-12}.
The dissociative photoionization cross sections of a SCH in~N$_2$ are known
from experiments at third-generation synchrotrons~\cite{Stolte:DP-98}.
From these we find~\cite{Buth:UA-12} for the fragment ion ratios
in~\eref{eq:fN2+} the
%
%
values~$f_{\mathrm{N}^+ + \mathrm{N}^+} = 0.74$
%
%
and $f_{\mathrm{N} + \mathrm{N}^{2+}} = 1 - f_{\mathrm{N}^+ + \mathrm{N}^+}
= 0.26$;
we use them to parametrize our rate equations to ensure that
the ion yields approach the values of SCH~decay from synchrotrons
in the limit of low \xray~intensities and fluences.
Auger and fluorescence decay widths and the corresponding
transition energies, electron binding energies,
and one-\xray-photon absorption cross sections
for multiply-ionized N~atoms can be found in~\cite{Buth:UA-12}
where they were determined with \emph{ab initio} computations.
The joint molecular and atomic rate equations form a system of
ordinary first-order linear differential equations which is solved
numerically with \textit{Mathematica}~\cite{Mathematica:pgm-V10.1,SuppData}
with the initial condition that N$_2$~is in its ground state, \ie,
$\tilde P_{223,223}(-\infty) = 1$, and all the other molecular states
and all atomic states are unpopulated, \ie, $\tilde P_{ijk,\ell mn}(-\infty)
= 0$ with~$0 \leq i,j,\ell,m \leq 2 \land 0 \leq k,n \leq 3 \land \lnot (
i=j=\ell=m=2 \land k=n=3)$ and $P_{ijk}(-\infty) = 0$ with~$0 \leq i,j \leq 2
\land 0 \leq k \leq 3$.

The ion yields of atomic fragments from molecules with a charge
of~$1 \leq j \leq 7$ are the molecular ion yields
\begin{eqnarray}
  \label{eq:yield}
  \widetilde{Y}_j = {P_j \over 1 - P_0} \; ,
\end{eqnarray}
which are experimentally mensurable quantities
and are defined as the renormalized charge-state probability where
\begin{eqnarray}
  P_j = \Sum_{\atopa{\scriptstyle \ell + m + n = 7 - j}{\scriptstyle
    0 \leq \ell, m \leq 2 \, \land \, 0 \leq n \leq 3}} P_{\ell mn}
    (\infty) \; ,
\end{eqnarray}
is the probability to find an atomic fragment with charge~$0 \leq j \leq 7$
for~$t \to \infty$ when molecular breakup is always assumed.
From the ion yields~\eref{eq:yield}, we calculate the average charge
state via
\begin{eqnarray}
  \label{eq:avrcharge}
  \bar q = \sum_{j = 1}^7 j \, \widetilde{Y}_j \; ;
\end{eqnarray}
it is an indicator of the amount of charge found on molecular fragments.

In the experiment, the beam line transmission is not
particularly well determined due to beam transport losses, which
leads to a large uncertainty in the pulse energy actually reaching
the sample;
we estimated that only $15 \%$--$35 \%$ of the nominal pulse energy
arrived at the sample~\cite{Hoener:FA-10,Buth:UA-12,Fang:MI-12}.
Therefore, the effective pulse energy~$E\X{P}$ and the rate of molecular
fragmentation~$\gamma$ [\Eref{eq:P212,223} and \eref{eq:P223}] in our
calculations are unknown and need to be found by comparing theoretical results
with experimental data.
For this purpose, we jointly determine~$E\X{P}$ and $\gamma$
by minimizing the criterion
\begin{eqnarray}
  \label{eq:criterion}
  C_j(E\X{P}, \gamma) = |\bar q\X{expt} - \bar q\X{theo}(E\X{P}, \gamma)|
    + 16 \> |\widetilde{Y}_{\mathrm{expt}, \, j}
    - \widetilde{Y}_{\mathrm{theo}, \, j}(E\X{P}, \gamma)| \; ;
\end{eqnarray}
it comprises the absolute value of the difference between
the experimental~$\bar q\X{expt}$ and the
theoretical~$\bar q\X{theo}(E\X{P}, \gamma)$
average charge state and the absolute value of 16~times the
difference between the experimental~$\widetilde{Y}_{\mathrm{expt}, \, j}$
and the theoretical~$\widetilde{Y}_{\mathrm{theo}, \, j}(E\X{P}, \gamma)$
ion yield of either~N$^+$ for~$j = 1$ or~N$^{2+}$ for~$j = 2$.
The second summand in~\eref{eq:criterion} causes the criterion to
have a fairly sharp global minimum where the prefactor of~16
is somewhat arbitrary but chosen large enough to ensure that the
second summand makes a large contribution thus effecting
a good agreement between experimental and theoretical average
charge states provided that simultaneously also the chosen
experimental and theoretical ion yield agree well;
only~N$^+$ or~N$^{2+}$~ion yields~\eref{eq:fN2+} are considered in the second
summand because only these are influence directly by~$\gamma$
[\Eref{eq:P212,223} and \eref{eq:P223}].
The ion yields from higher charge states are only indirectly
impacted by fragmentation~\cite{Buth:UA-12}.

\begin{table}
  \caption{Parameters of the molecular rate-equation model.
           The LCLS nominal FWHM pulse duration is~$\tau\X{X}$ and the
           effective pulse duration is~$\tau'\X{X}$~\cite{Buth:UA-12}.
           A nominal pulse energy of~$0.15 \U{mJ}$ is specified
           for $4 \U{fs}$~pulses and $0.26 \U{mJ}$ for the remaining
           three pulse durations of~$7 \U{fs}$, $80 \U{fs}$, and $280 \U{fs}$.
           The effective LCLS pulse energy from molecular rate equations
           is~$E\X{P}$, \ie, the energy which arrives at the sample.
           The fragmentation time for nominal pulse durations from an
           optimization~\eref{eq:criterion} with respect to the~N$^+$~yield
           is~$\gamma^{-1}_{\mathrm{N}^+}$ and with respect to
           the~N$^{2+}$~yield it is~$\gamma^{-1}_{\mathrm{N}^{2+}}$;
           correspondingly $\gamma^{\prime \, -1}_{\mathrm{N}^+}$
           and $\gamma^{\prime \, -1}_{\mathrm{N}^{2+}}$ are the fragmentation
           times for effective pulse durations.
           No fragmentation times could be determined for $4 \U{fs}$ and
           $7 \U{fs}$~pulses.
           The LCLS photon energy is~$1100 \eV$.}
  \begin{indented}
    \item[]\begin{tabular}{@{}lllllll}
      \br
       $\tau\X{X}$ [$\mathrm{fs}$] & $\tau'\X{X}$ [$\mathrm{fs}$] & $E\X{P}$
       [$\mathrm{mJ}$] & $\gamma^{-1}_{\mathrm{N}^+}$ [$\mathrm{fs}$] &
       $\gamma^{\prime \, -1}_{\mathrm{N}^+}$ [$\mathrm{fs}$] &
       $\gamma^{-1}_{\mathrm{N}^{2+}}$ [$\mathrm{fs}$] &
       $\gamma^{\prime \, -1}_{\mathrm{N}^{2+}}$ [$\mathrm{fs}$] \\
      \mr
                 280 &           112\phantom{.0} & $0.38 \times 0.26$
                     & 246 & 95 & 201 & 76 \\
       \phantom{0}80 & \phantom{0}40\phantom{.0} & $0.30 \times 0.26$
                     & \phantom{0}83 & 40 & 60 & 26 \\
       \phantom{00}7 & \phantom{00}2.8           & $0.16 \times 0.26$ & & \\
       \phantom{00}4 & \phantom{00}1.6           & $0.23 \times 0.15$ & & \\
      \br
    \end{tabular}
  \end{indented}
  \label{tab:parameters}
\end{table}

We list in~\tref{tab:parameters} the effective pulse energies
and fragmentation times obtained by minimizing the
criterion~\eref{eq:criterion} for $80 \U{fs}$ and
$280 \U{fs}$~nominal LCLS pulse durations.
It was found that the nominal pulse durations specified by the accelerator
electron beam parameters~\cite{Hoener:FA-10,Young:FE-10} are substantially
longer than the effective pulse durations
at the sample~\cite{Young:FE-10,Dusterer:FS-11} for which
values for the fragmentation time are also given.
As the dependence of $4 \U{fs}$ and $7 \U{fs}$~pulses on~$\gamma$
is very weak, it was not possible to extract it from~\eref{eq:criterion}.
In order to obtain the effective pulse energy for the short
pulse durations, $\gamma_{\mathrm{N}^+}$~is used with the value from
the nominal duration of the $280 \U{fs}$~pulses.
The percentages of the nominal pulse energies that effectively
arrive at the sample lie almost in the range
of~$15 \%$--$35 \%$ which was specified in the experiment~\cite{Hoener:FA-10}.
The percentages in~\tref{tab:parameters} of this work have a close agreement
with those of the fragmentation-matrix model in Table~I of~\cite{Buth:UA-12}
which are~$26 \%$, $16 \%$, $25 \%$, and $31 \%$ for pulse durations
of~$4 \U{fs}$, $7 \U{fs}$, $80 \U{fs}$ and $280 \U{fs}$, respectively.
This means that the effective energy of the LCLS pulses arriving at the sample
decreases with the shortening of the pulse.
A reanalysis of the experimental data of~\cite{Hoener:FA-10}
in~\cite{Fang:MI-12} confirmed the predicted drop in
the effective pulse energy.

Comparing~$\gamma^{-1}_{\mathrm{N}^+}$ and $\gamma^{-1}_{\mathrm{N}^{2+}}$
for a specific pulse duration in~\tref{tab:parameters} reveals a noticeable
difference which is due to the fact that different pathways contribute
to these ion yields.
For example, N$^+$~ions are not only produced by~\eref{eq:fN2+sym}
but, \eg, also by valence ionization of~N;
and N$^{2+}$~ions do not only result from~\eref{eq:fN2+asy} but, \eg,
also from Auger decay of a SCH in a N$^+$~ion.
As neither N$^+$ nor N$^{2+}$ are solely produced by molecular fragmentation,
this causes a systematic error, if we use~\eref{eq:criterion},
which can be estimated from the differences between the respective
fragmentation times in~\tref{tab:parameters} for the same pulse duration.
Another source of error are the variations of the fragmentation times
for a specific ion yield with the pulse duration, \eg,
$\gamma^{-1}_{\mathrm{N}^+}$, obtained from~$80 \U{fs}$ and $280 \U{fs}$~pulses.
This happens because the contribution of the pathways varies when the
pulse duration is shortened as the peak \xray~intensity increases.
The same holds true for the fragmentation times found from the effective
pulse durations which are not known particularly precisely.
In contrast to the theoretical results from the fragmentation-matrix
model~\cite{Buth:UA-12}, which were found to depend on the pulse
duration only weakly, for molecular rate equations the determination
of the fragmentation rate~$\gamma$ is sensitive to the actual pulse duration.
Overall, we find~$\gamma^{-1}$ to be in the range of~$40$--$250 \U{fs}$ which
is comparable to the value of~${\sim} 100 \U{fs}$ found
in~\cite{Beylerian:CE-04}.
The fragmentation time~$\gamma^{-1}$
%
%
is much longer than the lifetime of a SCH
%
%
of~$6.73 \U{fs}$~\cite{Buth:UA-12} but is faster than the time for SCH
production by photoionization for the~$280 \U{fs}$~pulse
%
%
which is~$348 \U{fs}$ at the peak intensity.
For the $4 \U{fs}$~short pulse, the lifetime of a SCH
is comparable with the time between two photoionizations
%
%
of~$10.1 \U{fs}$ whereas the time of molecular fragmentation~$\gamma^{-1}$
is about an order of magnitude longer~\cite{SuppData}.

\begin{figure}[tb]
  \begin{center}
    (a)~\includegraphics[clip,width=0.45\hsize]{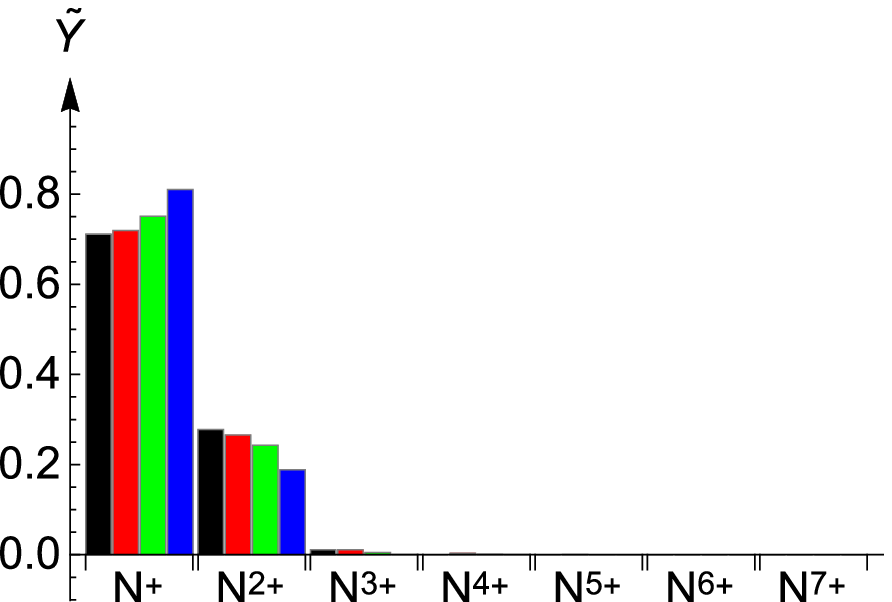}\hfill
    (b)~\includegraphics[clip,width=0.45\hsize]{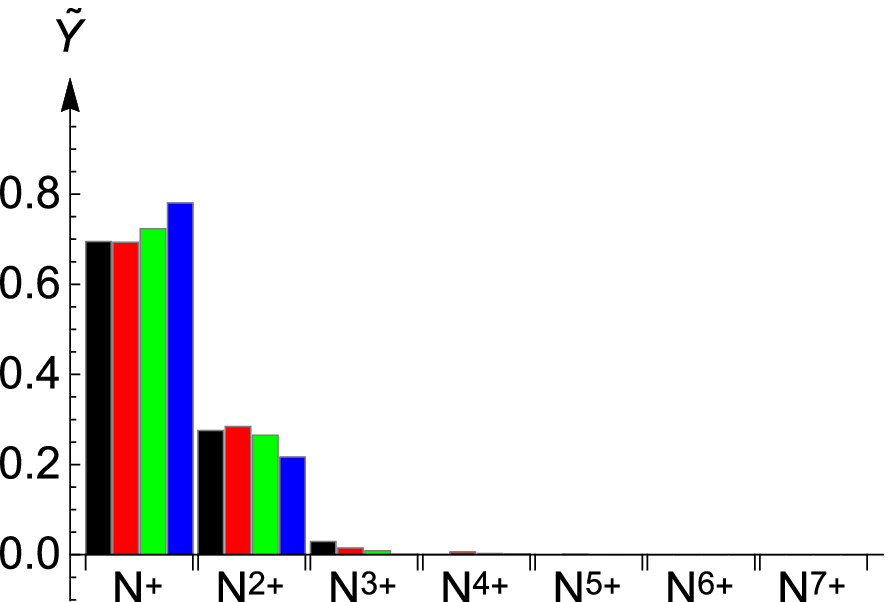}\linebreak
    (c)~\includegraphics[clip,width=0.45\hsize]{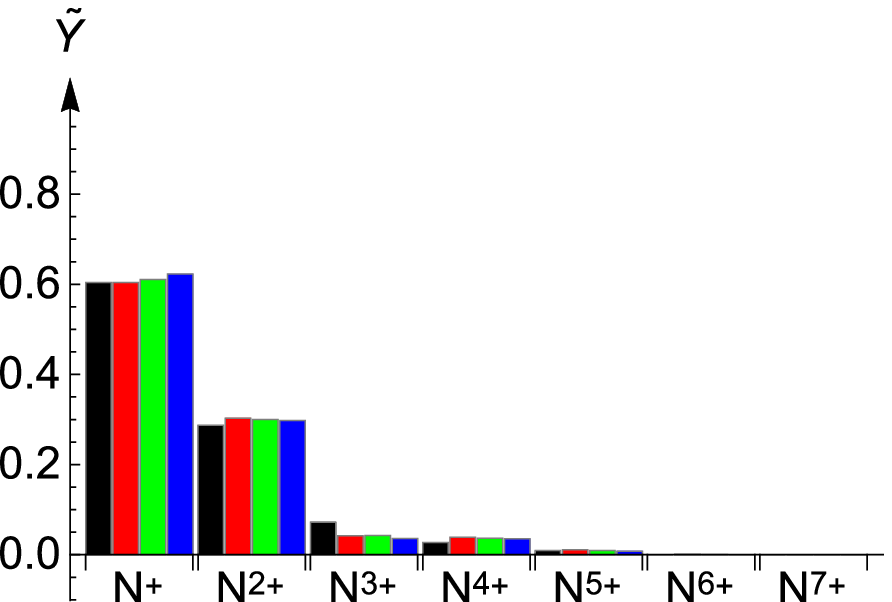}\hfill
    (d)~\includegraphics[clip,width=0.45\hsize]{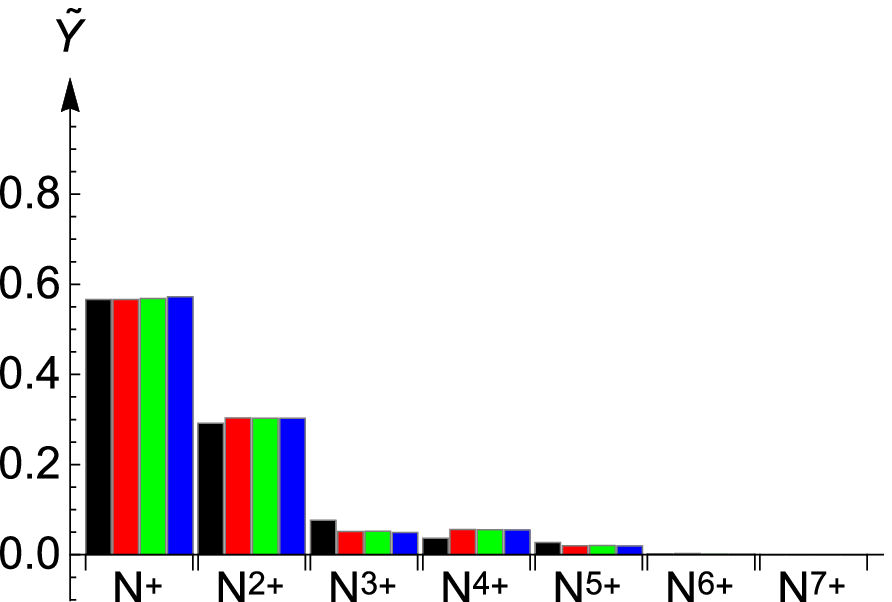}
    \caption{(Color) Ion yields of~N$_2$ ionized by LCLS x-rays with FWHM
             durations of (a)~$4 \U{fs}$, (b)~$7 \U{fs}$, (c)~$80 \U{fs}$,
             and (d)~$280 \U{fs}$.
             Experimental data (\textbf{\textcolor{mablack}{black}} bars);
             results of the molecular rate-equation model with all channels
             included (\textbf{\textcolor{mared}{red}} bars), only with SCH
             and tsDCH channels included (\textbf{\textcolor{magreen}{green}}
             bars), and only with SCH channels included
             (\textbf{\textcolor{mablue}{blue}} bars).
             The fragmentation time of the molecular dication~N$_2^{2+}$
             is~$\gamma^{-1} = 246 \U{fs}$ for all pulses but
             $80 \U{fs}$~pulses for which $\gamma^{-1} = 83 \U{fs}$ is used.
             Other parameters are taken from~\tref{tab:parameters}.}
    \label{fig:molionyldfrg}
  \end{center}
\end{figure}

The ion yields of~N$_2$ from our molecular rate-equation model and
the experiment are depicted in~\fref{fig:molionyldfrg} where
also theoretical ion yields are shown when certain DCH~channels
are closed.
The results from the molecular rate equations agree nicely
with the experimental data when all channels are included.
Considering the results for which the DCH-containing
channels are omitted, \ie, only the SCH~channel is included, we see that the
dominant features of the ion yields are reproduced.
Specifically, the influence of DCHs is small for~$280 \U{fs}$~pulses
but DCHs have a substantial influence on the ion yields
for the $4 \U{fs}$~pulse.
This result is consistent with the analysis in Section~V of~\cite{Buth:UA-12}.
The longer the pulse duration, the lower is the peak \xray~intensity
of the pulse.
Hence the rate of photoionization is fairly small for long pulses
and Auger decay and molecular fragmentation are the dominant
processes compared with DCH formation.
On the contrary, for short pulses, the molecules are ionized faster
resulting in DCHs but nuclear dynamics can be neglected
because of the short pulse duration and valence charges are
mostly shared equally between the two nitrogen atoms~\cite{Buth:UA-12}.
The artificial closing of DCH channels inhibits \xray~absorption and thus
leads to an underestimation of the amount of ionization which means that
the ion yields of low charge states in~\fref{fig:molionyldfrg} are higher
than the ones from the complete model and the higher charge states are
lower, respectively.

\begin{figure}[tb]
  \begin{center}
    (a)~\includegraphics[clip,width=0.45\hsize]{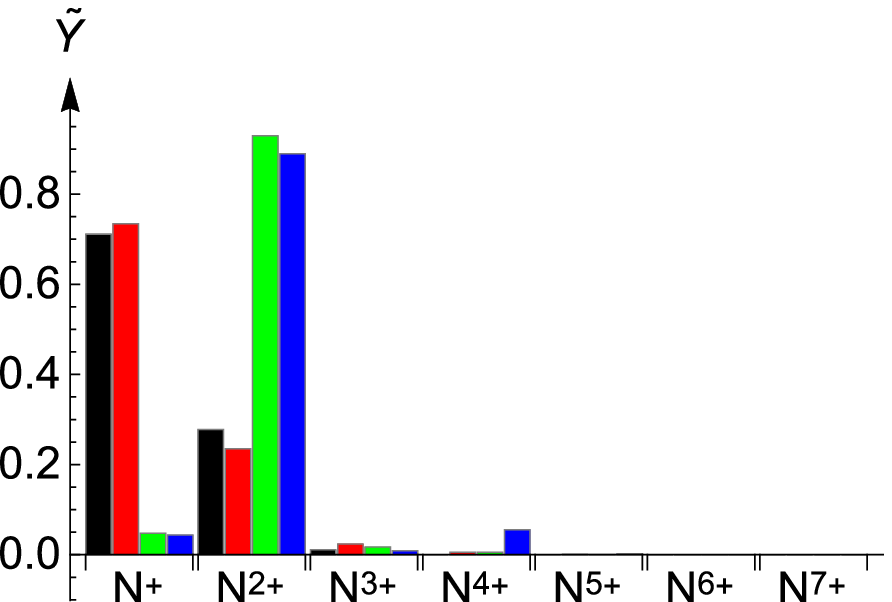}\hfill
    (b)~\includegraphics[clip,width=0.45\hsize]{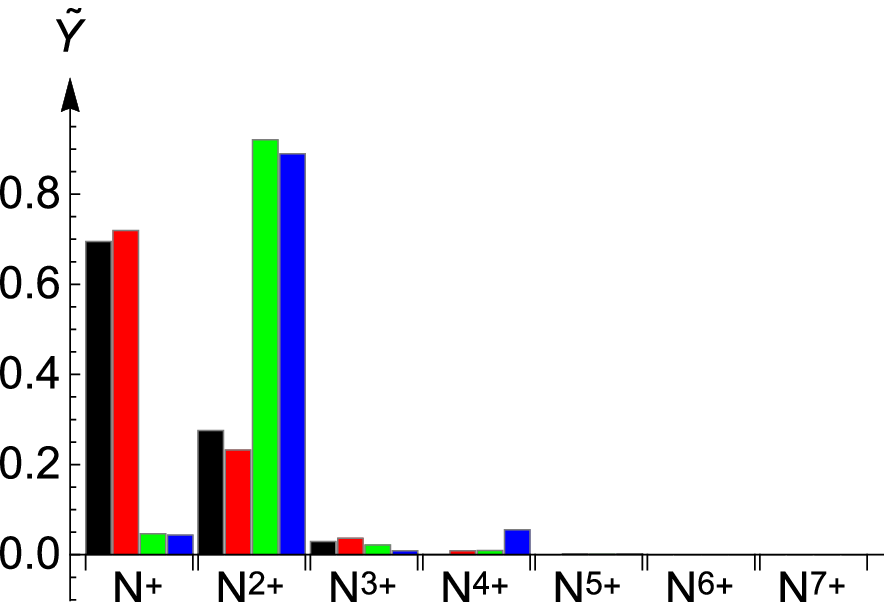}\linebreak
    (c)~\includegraphics[clip,width=0.45\hsize]{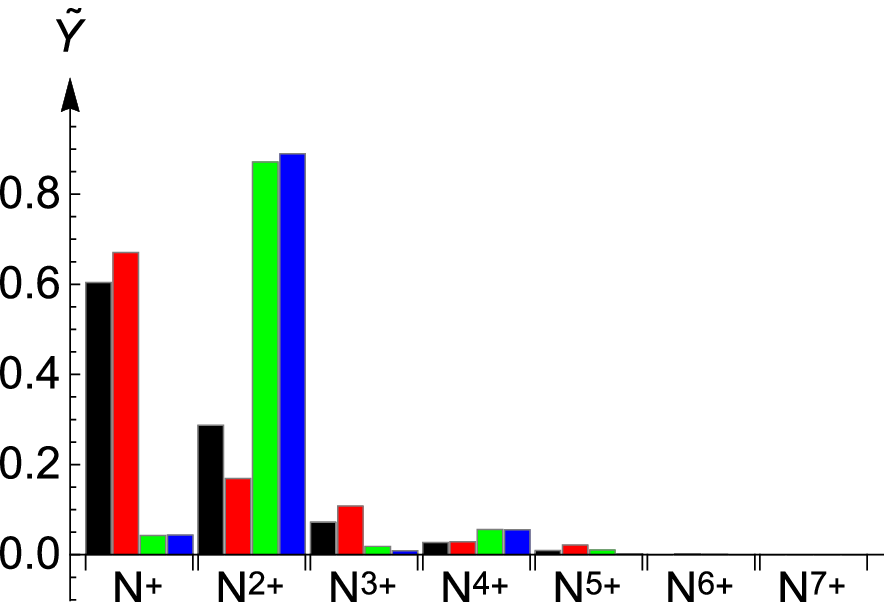}\hfill
    (d)~\includegraphics[clip,width=0.45\hsize]{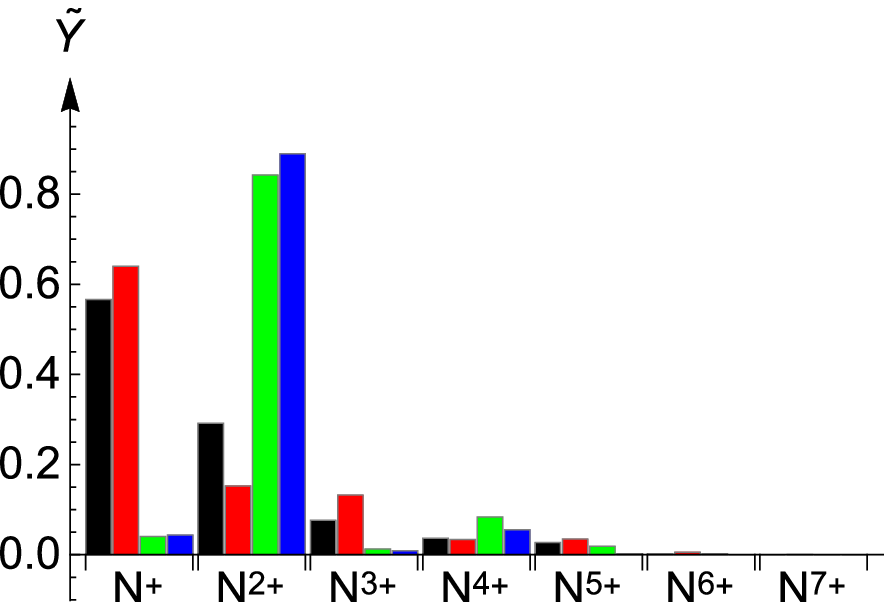}
    \caption{(Color) Ion yields of~N$_2$ ionized by LCLS x-rays with FWHM
             durations of (a)~$4 \U{fs}$, (b)~$7 \U{fs}$, (c)~$80 \U{fs}$, and
             (d)~$280 \U{fs}$.
             We display experimental data (\textbf{\textcolor{mablack}{black}}
             bars) alongside theoretical results of the molecular rate-equation
             model with all channels for~$\gamma \to \infty$
             (\textbf{\textcolor{mared}{red}} bars) and for~$\gamma = 0$
             (\textbf{\textcolor{magreen}{green}} bars).
             Further, we give theoretical results of the atomic rate-equation
             model (\textbf{\textcolor{mablue}{blue}} bars).
             Other parameters are taken from~\tref{tab:parameters}.}
    \label{fig:molionyldnofrg}
  \end{center}
\end{figure}

Merits of the molecular rate-equation model over the fragmentation
matrix model from~\cite{Buth:UA-12} are that the former takes explicitly
the time scale of molecular fragmentation into account and offers a
description of the dynamics of the population of charge states whereas
the latter provides only fragmentation constants.
In order to reveal the important role of the fragmentation rate
introduced in the molecular rate-equation model [\Eref{eq:P212,223}
and \eref{eq:P223}],
in~\fref{fig:molionyldnofrg}, a comparison of the ion yields is shown
which are obtained from our molecular rate-equation model in the two
extreme cases of~$\gamma \to \infty$ and $\gamma = 0$
together with the results of the single-atom model
and the experimental data.
When immediate fragmentation at the instant of the formation of
the metastable molecular dication~N$_2^{2+}$ is assumed
\ie, $\gamma \to \infty$, the molecular rate-equation model
still gives similar ion yields as the experimental data.
Specifically, the theoretical ion yields follow the same trend
as the experimental ones.
Yet we see clearly from~\fref{fig:molionyldnofrg}, that the rather
good agreement for~$4 \U{fs}$~pulses becomes progressively
worse for longer pulses.
This observation reveals the increasing influence of
the fragmentation time scale which eventually becomes comparable
with the FWHM duration of the long pulses.
As molecular breakup occurs immediately in~\eref{eq:P212,223} and
\eref{eq:P223} and the hierarchy of molecular rate equations is
quenched at~\eref{eq:P212,223} because all probability is funneled
into the atomic rate equation~\eref{eq:P223}, several of the molecular
rate equations for the trication are thus not used.
Conversely, the other limit of the fragmentation time scale
is~$\gamma = 0$ which means that fragmentation of the metastable
molecular dication~N$_2^{2+}$ is neglected.
One can see clearly from~\fref{fig:molionyldnofrg} the large
differences between the experimental results and the theoretical
calculations.
The ion yields in this limit agree well with the atomic
rate-equation model but poorly with the experimental data.
Specifically, the molecular rate equations in this limit do
not describe the experimental observation that the ion yields of the lowly
charged ions are higher than those of the highly charged ions
(see~\cite{Buth:UA-12} for details of the single-atom model).
We also observe in~\fref{fig:molionyldnofrg} that
the variation of the ion yields with the pulse duration
is substantially reduced for the atomic rate equations
which indicates that the heightened sensitivity to the pulse duration
is due to molecular fragmentation.
In other words, the metastability of the molecular dication~N$_2^{2+}$ plays an
essential role in the interaction of~N$_2$ with \xray~pulses
and its finite rate of dissociation must not be neglected.

\begin{figure}[tb]
  \begin{indented}
    \item[]\includegraphics[clip,width=0.8\hsize]{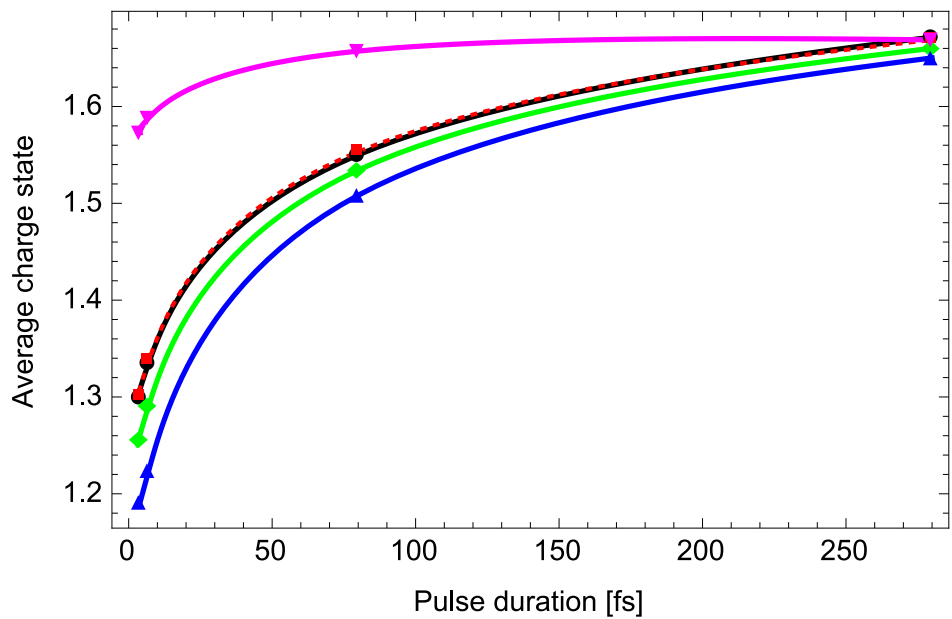}
  \end{indented}
  \caption{(Color) Average charge state~$\bar q$ [\Eref{eq:avrcharge}]
           of~N$_2$ ionized with LCLS~pulses of varying duration.
           The measured and computed data points are shown and
           connected by spline interpolation.
           The experimental~$\bar q$ is shown as
           \textbf{\textcolor{mablack}{black}} circles, the
           theoretical~$\bar q$, calculated with molecular rate
           equations is shown as \textbf{\textcolor{mared}{red}}
           squares (SCH, tsDCH, and ssDCH channels),
           \textbf{\textcolor{magreen}{green}}
           diamonds (only SCH and tsDCH channels),
           \textbf{\textcolor{mablue}{blue}} triangles up (only SCH channels),
           and \textbf{\textcolor{mamagenta}{magenta}} triangles down
           (constant pulse energy, SCH, tsDCH, and ssDCH channels).
           The parameters are the same as in~\fref{fig:molionyldfrg}.}
  \label{fig:molavchrg}
\end{figure}

In~\fref{fig:molavchrg}, we display the average charge state~$\bar q$
of~\eref{eq:avrcharge} obtained from the experiment~\cite{Hoener:FA-10}
and the molecular rate-equation model.
\Fref{fig:molavchrg} shows that the model reproduces very well the decreasing
tendency of~$\bar q$ with the decrease of the pulse duration.
This comes to us as no surprise because we fitted~\eref{eq:criterion}
the theoretical~$\bar q$ to the experimental value in order to determine
the effective pulse energy at the sample.
Yet the results for which certain DCH channels are closed provide
novel insights.
Inspecting~\fref{fig:molavchrg}, we find that the SCH~channel
determines the main character of the dependence of~$\bar q$
on the pulse duration.
It describes well the nuclear and electronic dynamics of~N$_2$
for long pulses.
That is because for, \eg, the $280 \U{fs}$~pulse, Auger decay is
the fastest process;
a core hole formed by photoionization of an inner-shell electron
is typically refilled by Auger decay before another core electron
is photoionized.
However, the restriction to only SCH~channels significantly
underestimates~$\bar q$ for short pulses compared with the experimental
result which can only be obtained by including the molecular
configurations with tsDCH and ssDCH.
This means that there is an appreciable probability of ionization of a
second core electron before Auger decay of the core hole occurs.
Furthermore, one can see from~\fref{fig:molionyldfrg} and
\fref{fig:molavchrg} that tsDCHs make a stronger contribution than ssDCHs
because of the larger cross section for the formation of a
tsDCH---as the \xray~photoabsorption cross section of an atom with
no core hole is larger than that with a single core hole---and
the lifetime of tsDCH is longer compared with the lifetime of
a ssDCH~\cite{Buth:UA-12}.

Overall, there is a drop in~$\bar q$ for short pulses compared with
long pulses in~\fref{fig:molavchrg}.
This \emph{observed} frustrated absorption is ascribed to three
effects:
first, the decrease in the \xray~absorption cross section due to SCH
and DCH formation, \ie, the effect of frustrated
absorption~\cite{Hoener:FA-10}\footnote{%
This is in contrast to~\cite{Buth:UA-12} where we focused on the sharing
of charges due to molecular breakup on page~12, left column, top.},
second, due to a predominantly symmetric sharing of the charges
for short pulses upon breakup as the molecular ion remains intact
during the interaction in contrast to long pulses with fragmentation
in the course of the interaction,
and, third, the variation in the pulse energy at the sample which drops
noticeably with shortening pulse duration [\Tref{tab:parameters}].
We also show~$\bar q$ in~\fref{fig:molavchrg} when the pulse energy is
constant.
In this case, the drop of~$\bar q$ is only caused by the change of
the breakup pattern and the effect of frustrated absorption.
Considering only~$4 \U{fs}$ and $7 \U{fs}$~pulses, there ought to be
no significant change of the breakup pattern as the time scale
for molecular dissociation of the metastable~N$_2^{2+}$ is
much longer than the pulse durations.
Nonetheless, there is a pronounced drop of~$\bar q$ for~$4 \U{fs}$~pulse
compared with~$7 \U{fs}$~pulses which we ascribe entirely to
frustrated absorption.

\Fref{fig:molavchrg} of this work corresponds to figure~4
of~\cite{Buth:UA-12} which shows the experimental~$\bar q$
together with $\bar q$~from the single-atom model that represents
an upper limit to~$\bar q$, the symmetric-sharing model which
provides a lower limit to~$\bar q$, and the fragmentation-matrix
model that reproduces the experimental~$\bar q$ very well for
short pulses but is less accurate for longer pulses as the
fragmentation time scale is not explicitly included in the model.
The agreement between the curves from the fragmentation-matrix model and
the molecular rate-equation model is not a coincidence;
it shows that the main physical effects are described.

Good agreement between theoretical calculation and experimental
measurement of the ion yields and the average charge state of~N$_2$
for LCLS pulses with varying durations was also obtained
in~\cite{Hoener:FA-10} saying that molecular rate equations
were used.
However, there, the molecular valence charge dynamics is modeled differently
to here because there N$_2^{4+}$~ions were assumed to fragment
into~$\mathrm{N}^{3+} + \mathrm{N}^{+}$ or $\mathrm{N}^{2+} + \mathrm{N}^{2+}$
with different ratios of~$\mathrm{N}^{3+} + \mathrm{N}^{+}$ versus
$\mathrm{N}^{2+} + \mathrm{N}^{2+}$---depending on whether
DCH~decay or two photoionization with Auger decay cycles~occurred---which
were claimed to be adjusted to obtain the best agreement with the
experimental data in the calculation there.
Furthermore, there the effective pulse energy at the sample was adjusted in the
range~$17 \%$--$21 \%$ of the nominal pulse energy---which was \emph{always}
[\tref{tab:parameters}] taken to be~$0.26 \U{mJ}$
in~\cite{Hoener:FA-10}---for pulses with FWHM~durations
of~$4 \U{fs}$, $7 \U{fs}$, $80 \U{fs}$, and $280 \U{fs}$.
Our computations reveal a stronger dependence of the average charge state
on the pulse energy in the range~$16 \%$--$38 \%$ of the nominal value
for the pulse energy of different durations where the nominal pulse energy
for~$4 \U{fs}$~pulses was~$0.15 \U{mJ}$ and $0.26 \U{mJ}$ otherwise
[\tref{tab:parameters}].
The differences between the model discussed here and the model
of~\cite{Hoener:FA-10} cannot be explained by the fact
that SASE-type pulses were used in~\cite{Hoener:FA-10} instead of
pulses with a Gaussian temporal shape employed here.
Namely, the analysis in~\cite{Buth:UA-12} reveals that the impact of
the spikiness of SASE pulses compared with a Gaussian pulse
is small and may be neglected in good approximation.
We believe that three aspects of our model cause the observed differences:
first, all possible one-photon absorption and decay channels of
an N~atom are determined from \emph{ab initio} computations~\cite{Buth:UA-12}
and are included in our model,
second, the molecular configurations up to~N$_2^{4+}$ are treated,
and, third, the crucial dynamics of dissociation of the
metastable molecular dication~N$_2^{2+}$ is considered by introducing
phenomenologically the fragmentation rate~$\gamma$.

\section{Conclusion}
\label{sec:conclusion}

We describe the interaction of intense and ultrafast x-rays with~N$_2$.
For this purpose, we lay the theoretical foundation for the description of
multi-\xray-photon absorption by molecules in rate-equation approximation
and investigate theoretically the quantum dynamics of~N$_2$ exposed to
\xray~pulses from LCLS.
The results are determined by the competition between photoabsorption,
decay processes, and molecular dissociation for
\xray~pulses of varying duration and similar nominal pulse energy.
Molecular configurations up to trications are included in the model
taking into account the fragmentation dynamics of the metastable
molecular dications and the redistribution of the valence electrons
between the two atomic nuclei upon breakup.
Atomic fragments after dissociation of the molecule are treated by atomic
rate equations, \ie, no molecular effects are regarded anymore.
The molecular rate-equation model describes the ionization dynamics
observed and allows us to calculate theoretical ion yields and the
average charge state which are then compared with the corresponding
experimental quantities.
Thereby, the effective pulse energy at the sample and the rate of fragmentation
of the molecular dication~N$_2^{2+}$ are determined from a comparison
of theoretical with experimental data.
We find a substantial progression in the effective pulse energy
which decreases pronouncedly from long to short pulses.
The ion yields and effective pulse energies at the sample from the
molecular rate equations agree well with those obtained previously
with the fragmentation-matrix model of~\cite{Buth:UA-12}.

In~\cite{Hoener:FA-10} frustrated absorption was \emph{observed}
as a drop of the average charge state for short pulses
compared with long pulses.
There the most pronounced impact on the average charge state
is due to the variation of the effective pulse energy as our
previous~\cite{Buth:UA-12} and present analysis reveals.
Furthermore, for short pulses the molecular ions remain intact
and fragment mostly in terms of symmetric sharing of charges
whereas for long pulses fragmentation occurs frequently prior
further absorption of x-rays leading to higher charge states in the latter
case compared with the former.
The effect of frustrated absorption describes the situation that
core-hole formation reduces the probability for further \xray~absorption.
It makes a significant contribution for short pulses compared with
long pulses as DCH channels are more important in the former.
Specifically, we find for long pulses with moderate \xray~intensity
that photoabsorption is the slowest process;
it is slower than the dissociation of the metastable dications~N$_2^{2+}$.
Molecular configurations with SCHs dominate where
Auger decay quickly refills SCHs and thus increase the chance
for further photoabsorption in the course of the interaction
with the pulse because the photoionization cross section of a filled
core shell is larger than the one of an only partially filled core shell.
Conversely, for short pulses with high \xray~intensities,
the rates of Auger decay and photoionization are comparable.
In this case, there is a much larger probability for the production
of~DCHs by sequential absorption of two x-rays within the lifetime
of the core hole.
Therefore, further \xray~absorption is reduced---compared with longer
pulses of the same effective pulse energy---causing a
smaller population of higher charge states.
Thus frustrated absorption results in a pronounced drop of the
average charge state in addition to the variation of the effective
pulse energy and the variation of fragmentation patterns.

We would like to point out that our molecular rate-equation model works
well for \xray~pulses at a photon energy of~$1100 \eV$.
We anticipate that the model remains applicable for lower photon
energies up to the point when the photon energy still exceeds the
highest ionization potential in the model of~$667.05 \eV$
for the ionization of the last $K$-shell electron of an otherwise
electron-bare nucleus~\cite{Buth:UA-12} as then resonances
still play no significant role.
Yet more experimental data are needed in order to assess whether
our model stays valid for photon energies higher than~$1100 \eV$
because then the single-electron response model may become insufficient
and multielectron effects such as shake off processes are
noticeable~\cite{Young:FE-10}.
Similar models to the ones discussed in~\cite{Buth:UA-12} and
here should be used to investigate other diatomic molecules in
order to find out how general our approach is and whether also
heteronuclear molecules are well described.
Also the role of interatomic electronic decay~\cite{Buth:IM-03}
ought to be investigated in the future.
We are confident that our theory can even be extended to
molecules with more than two atoms.

\begin{CJK*}{UTF8}{}
\ack
We are grateful to Mau Hsiung Chen ({\CJKfamily{bsmi}陳茂雄}),
Ryan N.{} Coffee, Li Fang ({\CJKfamily{gbsn}方力}),
Matthias Hoener, and Christoph H.{} Keitel for helpful discussions.
J-CL thanks for support by the National Science Foundation of China under
grant Nos.~11204078 and 11574082, and the Fundamental Research Funds
for the Central Universities of China under grant No.~2015MS54.
CB~was supported by the National Science Foundation under
grant~Nos.~PHY-0701372 and PHY-0449235 and by a Marie Curie International
Reintegration Grant within the 7$^{\mathrm{th}}$~European Community
Framework Program (call identifier: FP7-PEOPLE-2010-RG, proposal No.~266551).
NB is grateful for funding by the Office of Basic Energy Sciences,
Office of Science, U.S.~Department of Energy, under Contract
No.~DE-SC0012376.
Portions of this research were carried out at the Linac Coherent Light
Source~(LCLS) at SLAC National Accelerator Laboratory.
LCLS is an Office of Science User Facility operated for the
U.S.~Department of Energy Office of Science by Stanford University.
JPC~and JMG~were supported through both the LCLS and
The PULSE Institute for Ultrafast Energy Science at the
SLAC National Accelerator Laboratory by the U.S.~Department of Energy,
Office of Basic Energy Sciences.
\end{CJK*}

\end{document}